\documentclass{article}
\usepackage[utf8]{inputenc}

\usepackage{graphicx}
\usepackage{authblk}
\usepackage[utf8]{inputenc}
\usepackage{mathptmx}
\usepackage{epsfig,amsopn}
\usepackage{graphicx}
\usepackage{braket}
\usepackage{float}
\usepackage{enumerate}
\usepackage[normalem]{ulem}
\usepackage{todonotes}
\newcommand{\eref}[1]{Eq.~(\ref{#1})}%
\begin{document}

\title{Tail-behavior roadmap for sharp restart}

\author[1]{Iddo Eliazar \thanks{email: eliazar@tauex.tau.ac.il}}
\author[1]{Shlomi Reuveni \thanks{email: shlomire@tauex.tau.ac.il}}
\affil[1]{School of Chemistry, Center for the Physics and Chemistry of Living Systems, The Sackler Center for Computational Molecular and Materials Science, and The Mark Ratner Institute for Single Molecule Chemistry, Tel Aviv University, Tel Aviv 6997801, Israel}

\maketitle

\begin{abstract}

\noindent 
Many tasks are accomplished via random processes. The completion time of such a task can be profoundly affected by restart: the occasional resetting of the task’s underlying random process. Consequently, determining when restart will impede or expedite task completion is a subject of major importance. In recent years researchers explored this subject extensively, with main focus set on average behavior, i.e. on mean completion times. On the one hand, the mean approach asserts the centrality of ``sharp restart'' -- resetting with deterministic (fixed) timers. On the other hand, a significant drawback of the mean approach is that it provides no insight regarding tail behavior, i.e. the occurrence likelihood of extreme completion times. Addressing sharp restart, and shifting the focus from means to extremes, this paper establishes a comprehensive tail-behavior analysis of completion times. Employing the reliability-engineering notion of hazard rate, the analysis yields a set of universal results that determine -- from a tail-behavior perspective -- when sharp restart will impede or expedite task completion. The universal results are formulated in terms of explicit and highly applicable hazard-rate criteria. With the novel results at hand, a universal average-\&-tail classification manual for sharp restart is devised. The manual specifies general scenarios in which -- rather counter-intuitively -- sharp restart has an opposite effect on average behavior and on tail behavior: decreasing mean completion times while dramatically increasing the likelihood of extreme completion times; and, conversely, increasing mean completion times while dramatically decreasing the likelihood of extreme completion times.

\bigskip


\end{abstract}

\newpage

\section{Introduction}

Consider a task whose completion requires the execution of a certain underlying process. What is the effect of restart -- i.e. resetting the underlying process while it is running -- on the task's completion time? The answer to this question depends on the completion-time statistics. For example, if the underlying process is deterministic then the completion time is fixed, and hence: restart will always prolong completion. However, if the underlying process is stochastic then the completion time is a random variable, and matters become intricate \cite{FPUR}-\cite{review}: statistically, while restart can impede completion, it can also expedite completion. 

The fact that restart can affect completion times -- and in some cases significantly so -- has a host of important practical applications. Examples include: randomized computer algorithms \cite{CS1}-\cite{CS3}, e.g. simulated annealing \cite{CS4}; first-passage times of random motions \cite{RM1}-\cite{RM10}, e.g. Brownian motion \cite{Diff1}-\cite{Diff9}; target-search by agents \cite{Search1}-\cite{Search8}, e.g. animals foraging for food \cite{Foraging1,Foraging2}; and chemical reactions at the molecular level \cite{CR1,CR2}, e.g. enzymatic catalysis \cite{MM1}-\cite{MM4}. The ``tasks" in the above examples, as well as restart in these examples, are diverse. Indeed, a simulated-annealing program is reset by adding a line of code, while the enzymatic conversion of molecule A to molecule B is inherently subject to resetting as enzymes continuously bind and unbind their substrates. In all of the above-mentioned examples, it is vitally important to determine when restart will impede or expedite completion times. 

To determine the effect of restart studies have, by and large, focused on average behavior: comparing the mean completion time with restart to the mean completion time without restart. In general, a given restart protocol uses a stochastic timer to schedule the durations between its consecutive resetting epochs. Restart protocols with deterministic (i.e. fixed) timers -- termed, in short, \emph{sharp restart} -- where found to be central due to the following key result \cite{FPUR,CheSok,Diff5,Search5}: if a given restart protocol increases/decreases mean completion -- then there exists a sharp-restart protocol that increases/decreases mean completion at least as much.

Average-behavior analysis provides researchers with universal criteria that determine when restart will increase or decrease mean completion times. In particular, highly applicable universal criteria are available for Poissonian restart (where the stochastic timers are exponentially-distributed) \cite{FPUR,Search8,Foraging2,MM1,MM2,LTRT}, and for sharp restart \cite{MPSR1,MPSR2}. The drawback of average-behavior analysis is that it provides no insight regarding tail-behavior, i.e. the occurrence likelihood of extremely large completion times. Mode and median analyses of restart \cite{Belan} share the same drawback.

The difference between average-behavior and tail-behavior is profound. A system whose design is based on average-behavior analysis will perform well in `usual times'; yet it may very well fail when hit by an extreme event -- a, so called, `Black Swan' \cite{BS1}-\cite{BS5}. 
Financial crashes, extreme weather phenomena, extreme geological phenomena, and pandemics are vivid examples of `Black Swans'. To design a given system to withstand extreme events, a tail-behavior analysis is an absolute must. 


To illustrate `average-behavior vs. tail-behavior' consider a paradigmatic physical model in the restart literature \cite{Diff7,Diff8,Search8,Foraging2}: the first-passage time to a target of a particle that undergoes diffusion with drift over the real line. If the drift is towards the target, and if it is strong enough, then sharp restart will increase the mean first-passage time \cite{MPSR1}. However, for the same strong drift, numerical computations show that (see Fig. 1): sharp restart can dramatically decrease the probability that the first-passage time be large. Hence, the decision to apply or not to apply sharp restart depends on what matters most -- average-behavior or tail-behavior. To date, while universal criteria are available for the average-behavior perspective, there are no available universal criteria for the tail-behavior perspective.

Addressing sharp restart, and setting the goal of bridging the knowledge gap between means and extremes, this paper establishes a comprehensive tail-behavior analysis. Using the reliability-engineering notion of hazard rates, the analysis yields a tail-behavior `roadmap' for sharp restart: a set of universal hazard-rate criteria that determine when resetting will increase or decrease the likelihood of extreme completion times. These novel hazard-rate criteria are explicit, and are highly applicable.

\begin{figure}[t]
\centering
\includegraphics[width=9cm]{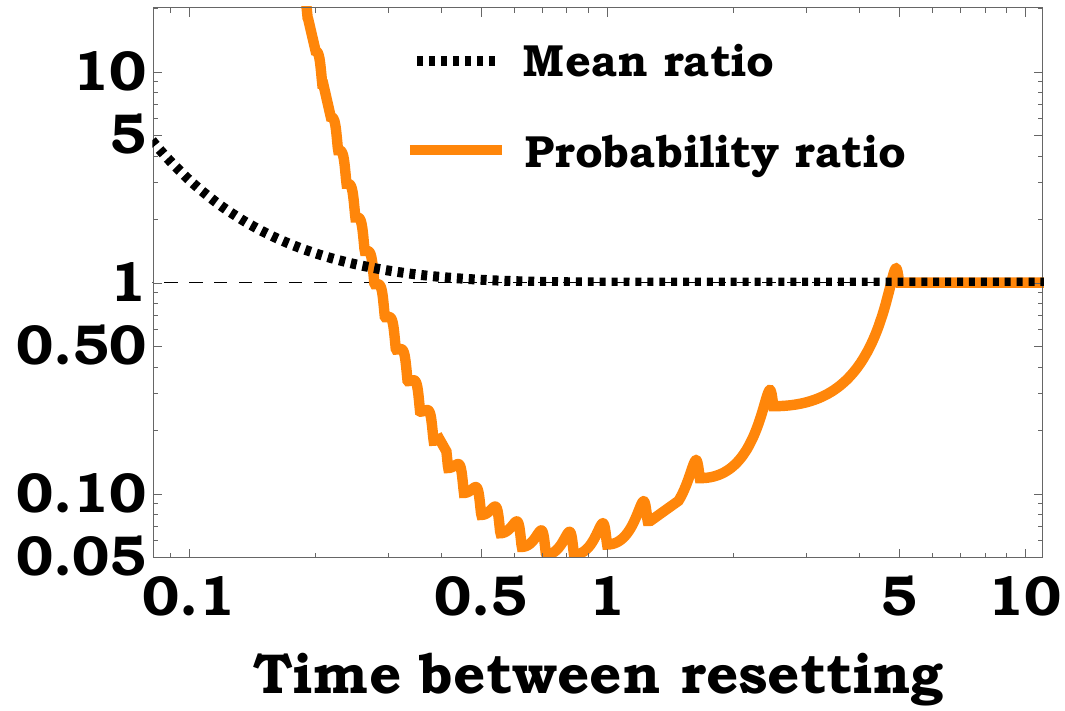}
\caption{Average-behavior vs. tail-behavior for the first-passage time (FPT) of diffusion with drift. Initiated from the origin, a particle moves randomly over the real line till it reaches a target positioned at the level $1$; the random motion has drift velocity $3$, and diffusion coefficient $1$. For this diffusion with drift the mean FPT is $1/3$, and the standard deviation of the FPT is $\sqrt{2/27}\simeq0.27$. Consequently, a FTP that is greater than $5$ time units is an `extreme event' whose occurrence probability is $\simeq5.55\times10^{-7}$. Applying sharp restart, and setting the horizontal axis to manifest the fixed duration between the sharp-restart consecutive resetting epochs: the dashed line depicts the ratio of the mean FPT under sharp restart to the `benchmark mean' $1/3$; and the solid orange line depicts the ratio of the FPT's extreme-event probability under sharp restart to the `benchmark probability' $\simeq5.55\times10^{-7}$. Evidently, sharp restart can reduce the extreme-event probability dramatically, while only barely increasing the mean FPT.}
\label{Intro_fig}
\end{figure}

Armed with the results established in this paper, we discover general scenarios in which the effect of sharp restart on average behavior and on tail behavior is antithetical. Specifically, general scenarios in which: sharp restart decreases mean completion times -- yet, simultaneously, it dramatically increases the likelihood of extreme completion times. And, conversely, general scenarios in which: sharp restart increases mean completion times -- yet, simultaneously, it dramatically decreases the likelihood of extreme completion times. These general scenarios underscore the indispensability of universal `extreme criteria'. 

The paper is organized as follows. After describing sharp restart as an algorithm that maps random inputs to random outputs (section 2), statistical formulations of the input-to-output map are presented (section 3), and the map's fixed points are explored (section 4). Then, the effect of sharp restart on
inputs with monotone increasing and monotone decreasing hazard rates is investigated (section 5), the asymptotic effect of sharp restart on general inputs is further investigated (section 6), and the general asymptotic results are discussed in detail (section 7). The paper concludes with a summary of the universal tail-behavior results established here, and with a manual for the universal classification of sharp restart from a joint average-behavior and tail-behavior perspectives (section 8). In particular, the manual specifies the general scenarios in which sharp restart has an opposite effect on average behavior and on tail behavior. The derivations of this paper's key results are detailed in the Methods (section 9).

\section{Sharp restart}

We consider a general task with completion time $T$, a positive-valued
random variable. To this task we apply restart with a deterministic timer $%
\tau $, a positive parameter. Specifically, we operate according to the following
three-steps \emph{sharp-restart algorithm}. Step I: initiate simultaneously the task and the timer. Step II: if the task is
accomplished up to the timer's expiration -- i.e. if $T\leq \tau$ -- then stop upon completion. Step III: if
the task is not accomplished up to the timer's expiration -- i.e. if $T>\tau$ -- then, as the timer expires, go back to Step I. 

The sharp-restart algorithm generates an iterative process of independent and statistically identical task-completion trials.\footnote{We note that this process appears also in the context of preventive maintenance \cite{BH}-\cite{BP2}.} This process halts during its first successful trial, and we denote by $T_{R}$ its halting time. Namely, $T_{R}$ is the overall time it takes -- when the sharp-restart algorithm is applied -- to complete the task. The algorithm is a non-linear mapping whose \emph{input} is the random variable $T$, whose \emph{output} is the random variable $T_{R}$, and whose (single) parameter
is the deterministic timer $\tau $.

Stochastically, the \emph{input-to-output map} $T\mapsto T_{R}$ is described as follows: 
\begin{equation}
T_{R}=\left\{ 
\begin{array}{lll}
T & & if\ T\leq \tau ,\\ 
& \ &  \\ 
\tau +T_{R}^{\prime } &  & if\ T>\tau ,%
\end{array}%
\right.   \label{21}
\end{equation}%
where $T_{R}^{\prime }$ is a copy of the random variable $T_{R}$ that
is independent of the random variable $T$. The top line on the right-hand
side of Eq. (\ref{21}) corresponds to the Step-II scenario of the
sharp-restart algorithm, and the bottom line corresponds to the Step-III
scenario. Indeed, if the Step-III scenario occurs then, as the timer expires, the task-completion process is restarted anew; the random variable $T_{R}^{\prime }$ is the halting time of the restarted process. 

Henceforth, we set the sharp-restart algorithm to initiate at time $t=0$, and thus the process of task-completion trials takes place over the non-negative time axis $t\geq 0$. Along this paper we use the following periodic parameterization of the time axis: $t=\tau n+u$, where $n=0,1,2,\cdots $, and where $0\leq u<\tau $. In this parameterization the timer $\tau $ is the underpinning period; $n=\lfloor t/\tau \rfloor $ is the floor of $t/\tau $; and $u=t-\tau n$ is the reminder of $t$ after its division by $\tau $. 

With regard to the process of task-completion trials, the periodic parameterization $t=\tau n+u$ has the following interpretation. If the halting time $T_{R}$ is realized at time epoch $t$, i.e. if $T_{R}=t$, then: $n$ is the number of unsuccessful trials; and $u$ is the time epoch, within the first successful trial, at which the task-completion process halted. 

\section{Statistical formulations}

There are alternative ways of characterizing the input's and output's statistical distributions. In this section we employ three such ways -- survival functions, density functions, and hazard functions -- to statistically formulate the input-to-output map $T\mapsto T_{R}$. Hazard functions, also known as \textquotedblleft hazard rates\textquotedblright and \textquotedblleft failure rates\textquotedblright, are widely applied in reliability engineering \cite{BP2}-\cite{Dhi}. As we shall see, hazard functions will turn out to be remarkably useful in the tail-behavior analysis of the sharp-restart algorithm.

Consider the input's and output's survival functions: $\bar{F}\left(t\right) =\Pr \left( T>t\right)$ and $\bar{F}_{R}\left( t\right) =\Pr \left(T_{R}>t\right)$; these terms manifest, respectively, the probabilities that the input $T$ and the output $T_{R}$ are not realized by time $t$. From a survival-function perspective, the input-to-output map $T\mapsto T_{R}$ is manifested by
\begin{equation}
\bar{F}_{R}\left( \tau n+u\right) =\bar{F}\left( \tau \right) ^{n}\bar{F}%
\left( u\right) .  \label{31}
\end{equation}%
Eq. (\ref{31}) appeared in \cite{BP2} (in the context of preventive maintenance), and it is explained as follows.

The output $T_{R}$ is not realized by time $t=\tau n+u$ if and only if two
events occur. Event $A$: the first $n$ task-completion trials are
unsuccessful. Event $B$: the task is not completed during the first $u$ time
units of the task-completion trial $n+1$. The probability that
a task-completion trial fails is $\Pr \left( T>\tau \right) =\bar{F}\left(
\tau \right) $, and the probability of event $B$ is $\Pr \left( T>u\right) =%
\bar{F}\left( u\right) $. As the task-completion trials are independent of
each other, the probability of the event $A$ is $\bar{F}\left( \tau \right)
^{n}$, and the probability of the event $A\cap B$ is $\bar{F}\left( \tau
\right) ^{n}\cdot \bar{F}\left( u\right) $. Hence, Eq. (\ref{31}) is obtined.

The input's and output's density functions are the negative derivatives of their survival functions: $f\left( t\right) =-\bar{F}^{\prime }\left(t\right)$ and $f_{R}\left( t\right) =-\bar{F}_{R}^{\prime }\left( t\right)$; these terms manifest, respectively, the likelihoods that the input $T$ and the output $T_{R}$ be realized at time $t$. Differentiating Eq. (\ref{31}) with respect to the variable $u$ yields the following density-function formulation of the input-to-output
map $T\mapsto T_{R}$:
\begin{equation}
f_{R}\left( \tau n+u\right) =\bar{F}\left( \tau \right) ^{n}f\left( u\right) 
.  \label{32}
\end{equation}%

The input's and output's hazard functions are the ratios of their density functions to their survival functions: $H\left( t\right) = f\left( t\right)/\bar{F}\left( t\right)$ and $H_{R}\left( t\right) = f_{R}\left( t\right)/\bar{F}_{R}\left(t\right)$.\footnote{Alternatively, the input's and output's hazard functions are the negative logarithmic derivatives of their survival functions: $H\left( t\right) = -\{ln[\bar{F}(t)]\}'$ and $H_{R}\left( t\right) = -\{ln[\bar{F}_R(t)]\}'$.} The terms $H\left( t\right)$ and $H_{R}\left( t\right)$ manifest, respectively, the likelihoods that the input $T$ and the output $T_{R}$ be realized at time $t$ -- provided the information that $T$ and $T_{R}$ were not realized up to time $t$. Dividing the sides of Eq.  (\ref{32}) by the corresponding sides of Eq. (\ref{31}) yields the following hazard-function formulation of the input-to-output map $T\mapsto T_{R}$:
\begin{equation}
H_{R}\left( \tau n+u\right) =H\left( u\right) .  \label{41}
\end{equation}

Eqs. (\ref{31})-(\ref{41}) provide different -- yet equivalent -- statistical formulations of the input-to-output map $T\mapsto T_{R}$. Indeed, in terms of their hazard functions, the input's and output's survival functions are given by $\bar{F}\left( t\right) =\exp \{-\int_{0}^{t}H(s)ds\}$ and $\bar{F}_{R}\left( t\right) =\exp \{-\int_{0}^{t}H_{R}(s)ds\}$ \cite{BP2}-\cite{Dhi}. The hazard functions offer, via Eq. (\ref{41}), a most compact and neat formulation of this map. From a hazard-function perspective, the sharp-restart algorithm is described as follows: it takes the input's hazard function over the temporal interval $ 0 \leq t< \tau $, and it generates from this segment -- via periodic repetition -- the output's hazard function (see Fig. 2).

\begin{figure}[t]
\centering
\includegraphics[width=9cm]{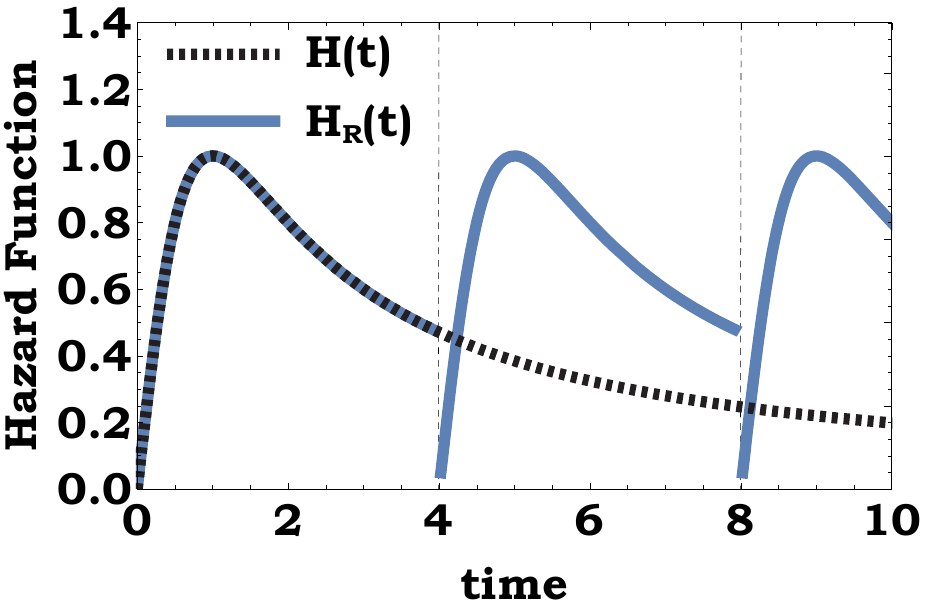}
\caption{Illustration of Eq. (\ref{41}), the hazard-function formulation of the input-to-output map $T\mapsto T_{R}$. Eq. (\ref{41}) is demonstrated via the example of a type-III Pareto input. The Pareto distributions, which comprise of four types, are the principal models of statistical power-laws in science and engineering \cite{Par}-\cite{Arn}. The type-III Pareto input is characterized by the survival function $\bar{F}\left( t\right)=1/(1+t^{p})$, as well as by the hazard function $H(t)=pt^{p-1}/(1+t^{p})$, where $p$ is a positive power. Here, for the Pareto power $p=2$, we plot the input's hazard function in dashed black. Also, for the timer parameter $\tau=4$, we plot the output's hazard function in solid blue. Note that the solid blue curve is produced by taking the temporal segment $0\leq t<4 $ of the dashed black curve, and by repeating it periodically.}
\label{Hazard_fig}
\end{figure}

\section{Fixed points}

The \emph{fixed points} of the input-to-output map $T\mapsto T_{R}$ are inputs that are statistically invariant to the action of this map. Namely, inputs $T$ such that $T_{R}=T$, the equality being in law. We now set the focus on these fixed points. 

From a hazard-function perspective the fixed points are characterized by $H_R(t)=H(t)$ ($t\geq 0$). Consequently, using  Eq. (\ref{41}), an input $T$ is a fixed point of the input-to-output map $T\mapsto T_{R}$ if and only if:
\begin{equation}
H\left( \tau n+u\right)=H\left( u\right) ,  \label{42}
\end{equation}
for all $n=0,1,2,\cdots $, and all $0\leq u<\tau $.

There are two types of fixed points: \emph{specific} and \emph{general}. A specific fixed point is with
respect to a specific timer $\tau $. For a specific timer $\tau $ it is evident from Eq. (\ref{42}) that: the 
\emph{specific fixed points} of the input-to-output map $T\mapsto T_{R}$ are inputs that are characterized by periodic
hazard functions with period $\tau $.  

A general fixed point is with respect to all timers $\tau $ \emph{simultaneously}. Eq. (\ref{42}) holds for all timers $%
\tau $ simultaneously if and only if the hazard function is constant. In turn, constant
hazard functions characterize Exponentially-distributed inputs \cite{BB}. Indeed, for a positive parameter $\lambda $ we have: $%
H(t)=\lambda $ ($t\geq 0$) if and only if $\bar{F}\left( t\right) =\exp
\left( -\lambda t\right) $ ($t\geq 0$). Hence, we assert that: the \emph{%
general fixed points} of the input-to-output map $T\mapsto T_{R}$ are Exponentially-distributed inputs. 

Exponentially-distributed inputs are characterized by the \emph{memoryless property} \cite{BB}:  $Pr(T>t+s|T>t)=Pr(T>s)$, for all $t \geq 0$ and $s\geq0$. It is evident from the memoryless property that applying the sharp-restart algorithm to Exponentially-distributed inputs will have no effect whatsoever on task-completion. Thus, the fact that Exponentially-distributed inputs are general fixed points of the input-to-output map $T\mapsto T_{R}$ follows also from the memoryless property.

\section{\label{5} Stochastic dominance}

Reliability engineering distinguishes two important classes of inputs \cite{BP2}-\cite{Dhi}: \emph{increasing
failure rate} (IFR), and \emph{decreasing failure rate} (DFR). The IFR and
DFR classes constitute, respectively, all inputs whose hazard functions are
monotone increasing and monotone decreasing. In this section we examine the effect of the input-to-output map $T\mapsto T_{R}$ on these classes of inputs.  

The IFR class manifests the following statistical behavior: the longer we wait for
an input $T$ to be realized -- the greater the likelihood that it will soon
be realized. The lifespans of aging systems -- e.g. cars, planes, machines,
and our own adult bodies -- are considered IFR. Namely, in aging systems the likelihood of
system-failure grows as the age of the system grows.

The DFR class manifests a statistical behavior that is antithetical to that of the IFR class. Specifically, for the DFR class: the longer
we wait for an input $T$ to be realized -- the smaller the likelihood that it will soon be realized. The lifespans of technologies -- e.g. the English
alphabet, the Gregorian calendar, the wheel, and the cutlery we use -- are considered DFR \cite{Tal}-\cite{Lind}. Indeed, the longer we have been using a technology, the more likely it is that we will keep on using it.

\begin{figure}[t]
\centering
\includegraphics[width=9cm]{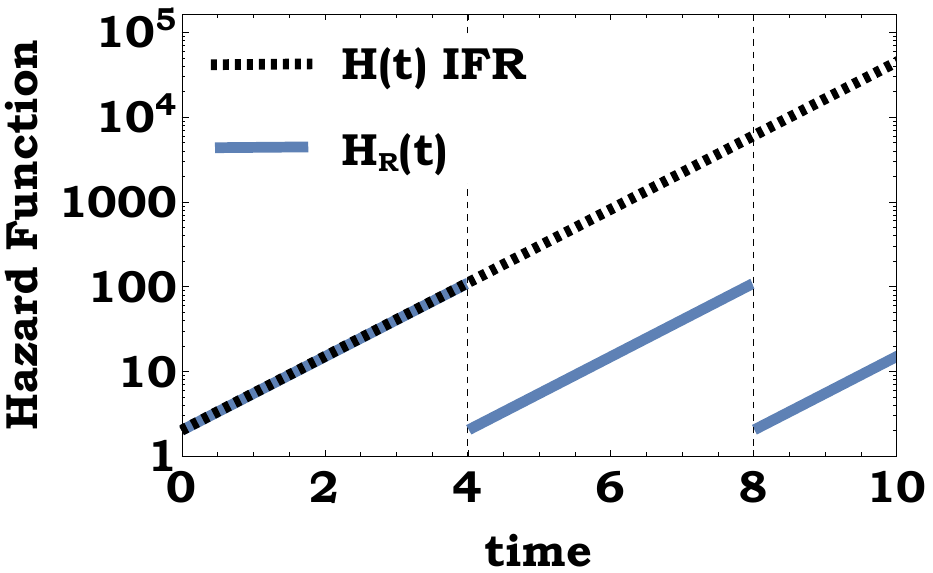}
\caption{Illustration of Eq. (\ref{41}) -- the hazard-function formulation of the input-to-output map $T\mapsto T_{R}$ -- in the case of IFR inputs. The IFR case is demonstrated via the example of a Gompertz input. The Gompertz distribution serves as a principal statistical model, in demography and in actuary, for adults' lifespans \cite{Gom}-\cite{PHG}; this distribution is generated by accelerating-change processes \cite{AccCha}, and is intimately related to Moore's law \cite{MooClo}. The Gompertz input is characterized by the survival function $\bar{F}\left( t\right) =\exp \left\{ -p\left[ \exp \left(t\right) -1\right] \right\} $, as well as by the monotone increasing hazard function $H\left( t\right) =p\exp \left( t\right) $, where $p$ is a positive parameter. Here, for the Gompertz parameter $p=2$, we plot the input's hazard function in dashed black. Also, for the timer parameter $\tau=4$, we plot the output's hazard function in solid blue. Note that, over the temporal ray $4<t<\infty$, the solid blue curve is strictly below the dashed black curve.}
\label{Hazard_fig}
\end{figure}

For general inputs Eq. (\ref{41}) implies that $H_{R}\left( t\right) = H\left( t\right) $ for all $t \leq \tau $. For IFR and DFR inputs Eq. (\ref{41}) further yields the following pair of observations. If the input is IFR then $H_{R}\left( t\right) <H\left( t\right) $ for all $t>\tau $ (see Fig. 3). And, if the input is DFR then $H_{R}\left( t\right) >H\left( t\right) $ for all $t>\tau $ (see Fig. 4). 
As noted above, in terms of their hazard functions, the input's and output's survival functions are given by $\bar{F}\left( t\right) =\exp \{-\int_{0}^{t}H(s)ds\}$ and $\bar{F}_{R}\left( t\right) =\exp \{-\int_{0}^{t}H_{R}(s)ds\}$. Also, in terms of their survival functions, the input's and output's means are given by $\mathbf{E}\left[ T%
\right] =\int_{0}^{\infty }\bar{F}\left( t\right) dt$ and $\mathbf{E}\left[
T_{R}\right] =\int_{0}^{\infty }\bar{F}_{R}\left( t\right) dt$. These survival-function formulae and mean formulae, combined together with the above IFR and DFR observations, yield the following pair of IFR and DFR results.

\begin{figure}[t]
\centering
\includegraphics[width=9cm]{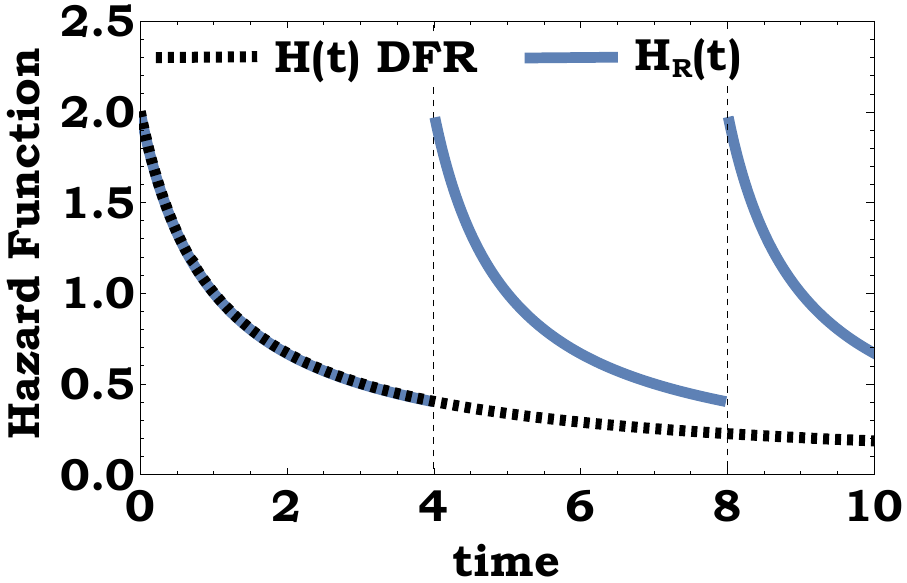}
\caption{Illustration of Eq. (\ref{41}) -- the hazard-function formulation of the input-to-output map $T\mapsto T_{R}$ -- in the case of DFR inputs. The DFR case is demonstrated via the example of a type-II Pareto input. As noted above, the Pareto distributions are the principal models of statistical power-laws in science and engineering \cite{Par}-\cite{Arn}. The type-II Pareto input is characterized by the survival function $\bar{F} \left( t\right) =1/(1+t)^{p}$, as well as by the monotone decreasing hazard function $H\left( t\right) =p/(1+t)$, where $p$ is a positive power. Here, for the Pareto power $p=2$, we plot the input's hazard function in dashed black. Also, for the timer parameter $\tau=4$, we plot the output's hazard function in solid blue. Note that, over the temporal ray $4<t<\infty$, the solid blue curve is strictly above the dashed black curve.}
\label{Hazard_fig}
\end{figure}

\begin{enumerate}
\item[$\bullet $] If the input is IFR then the output's survival function is larger than the
input's survival function 
\begin{equation}
\bar{F}_{R}\left( t\right) >\bar{F}\left( t\right) \label{51}
\end{equation}%
for all $t> \tau $; consequently, the output's mean is larger than the
input's mean, $\mathbf{E}\left[ T_{R}\right] >\mathbf{E}\left[ T\right] $.

\item[$\bullet $] If the input is DFR then the output's survival function is smaller than the
input's survival function
\begin{equation}
\bar{F}_{R}\left( t\right) <\bar{F}\left( t\right) \label{52}
\end{equation}%
for all $t> \tau $; consequently, the output's mean is smaller than the
input's mean, $\mathbf{E}\left[ T_{R}\right] <\mathbf{E}\left[ T\right] $.
\end{enumerate}

From a survival-function perspective, as well as from a mean perspective,
these results assert that: the sharp-restart algorithm impedes
task-completion in the case of IFR inputs, and expedites task-completion in
the case of DFR inputs. The IFR and DFR results hold valid for all timers $%
\tau $ simultaneously. Eqs. (\ref{51})-(\ref{52}) manifest \emph{stochastic dominance} \cite{SDN}-\cite{Levy}: that of the output $T_{R}$ over the input $T$ [Eq. (\ref{51})], and that of the input $T$ over the output $T_{R}$ [Eq. (\ref{52})].

\section{\label{6} Asymptotic stochastic dominance}

The IFR and DFR results of the previous section enable an immediate determination of the
impeding/expediting effect of  sharp restart on
task-completion. However, these results come with a caveat: they are not
always applicable. Indeed, while many inputs are IFR (e.g. the Gompertz input of Fig. 3), and while many other inputs are DFR (e.g. the type-II Pareto input of Fig. 4), there are also many inputs that are neither IFR nor DFR (e.g. the type-III Pareto input\footnote{The hazard function of the type-III Pareto input, in the parameter range $p>1$, has a unimodal shape.} of Fig. 2). Can we, by modifying the setting underpinning the IFR and DFR results, obtain
results that are applicable to \emph{all} inputs? The answer, as we shall argue and establish in this section, is affirmative.

The IFR and DFR results of section \ref{5} focus on the input's and output's survival
functions, $\bar{F}\left( t\right) $ and $\bar{F}_{R}\left( t\right) $, over
the temporal ray $ \tau < t < \infty $. We now shift the focus from the
temporal ray $ \tau < t < \infty $ to the temporal limit $t\rightarrow \infty $.
Specifically, we now set the focus on the \emph{asymptotic tail-behavior},
relative to each other, of the input's and output's survival functions.

To that end we use two `end terms' of the input's hazard function:
zero-end and infinity-end. The zero-end term is the average 
\begin{equation}
\bar{H}\left(\tau \right) =\frac{1}{\tau }\int_{0}^{\tau }H\left( t\right) dt
\end{equation}
of the
input's hazard function over the temporal interval $ 0 \leq t < \tau $; in terms of the input's survival function, this average is given by $\bar{H}(\tau) =-\frac{1}{\tau } \ln{[\bar{F}(\tau)]}$. The
infinity-end term is the limit 
\begin{equation}
H\left( \infty \right) =\lim_{t\rightarrow \infty }H\left( t\right)
\end{equation}
of the input's hazard function at infinity; we assume that this limit exists in the wide sense, i.e. $0\leq H\left( \infty
\right) \leq \infty $.

On the one hand, the survival-function formula $\bar{F}\left( t\right) =\exp \{-\int_{0}^{t}H(s)ds\}$ implies that the limit $H\left( \infty \right) $ affects the asymptotic tail-behavior of the input's survival function $\bar{F}\left(t\right)$. On the other hand, the survival-function formula  $\bar{F}_{R}\left( t\right) =\exp \{-\int_{0}^{t}H_{R}(s)ds\}$ together with Eq. (\ref{41}) imply that the average $\bar{H}\left( \tau \right)$ affects the asymptotic tail-behavior of the output's survival function $ \bar{F}_{R}\left( t\right)$. In turn, we find that the relative asymptotic tail-behavior of the input's and output's survival functions is determined by the difference between the limit $H\left( \infty \right) $ and the average $\bar{H}\left( \tau \right)$ as follows:

\begin{equation}
\lim_{t\rightarrow \infty }\frac{1}{t}\ln \left[ \frac{\bar{F}_{R}\left(
t\right) }{\bar{F}\left( t\right) }\right] =H\left( \infty \right) -\bar{H}%
\left( \tau \right) .  \label{60}
\end{equation}%
The proof of Eq. (\ref{60}) is detailed in the Methods. 

When $ H\left( \infty \right) <\infty $, an alternative way of formulating Eq. (\ref{60}) is: $\bar{F}_{R}\left(
t\right)/\bar{F}\left( t\right)  =\exp\left\{t[H\left( \infty \right) -\bar{H}%
\left( \tau \right) + \delta(t)]\right\}, \label{exp_approx}$
where $\delta(t)$ is a temporal function that vanishes at infinity, $\lim_{t\to\infty} \delta(t)=0$. As explained in the Methods, Eq. (\ref{60}) yields the following pair of asymptotic results.

\begin{enumerate}
\item[$\bullet $] If $\bar{H}\left( \tau \right) <H\left( \infty \right) $
then the output's survival function decays infinitely slower than the
input's survival function:%
\begin{equation}
\lim_{t\rightarrow \infty }\frac{\bar{F}_{R}\left( t\right) }{\bar{F}\left(
t\right) }=\infty .  \label{61}
\end{equation}

\item[$\bullet $] If $\bar{H}\left( \tau \right) >H\left( \infty \right) $
then the output's survival function decays infinitely faster than the
input's survival function:%
\begin{equation}
\lim_{t\rightarrow \infty }\frac{\bar{F}_{R}\left( t\right) }{\bar{F}\left(
t\right) }=0.  \label{62}
\end{equation}
\end{enumerate}

From an asymptotic tail-behavior perspective these results assert when sharp restart dramatically impedes task-completion, and
when it dramatically expedites task-completion. Eq. (\ref{61}) and Eq. (\ref{62}) are, respectively, the ``asymptotic stochastic dominance'' counterparts of Eq. (\ref{51}) and Eq. (\ref{52}).

Last, we note that the asymptotic results of Eq. (\ref{61}) and Eq. (\ref{62}) are in full accord, respectively, with the IFR and DFR results of section \ref{5}. Indeed, if the input is IFR then its hazard function is monotone increasing, hence $\bar{H}\left( \tau \right) < H\left( \infty \right) $ for all timers $\tau $, and thus we conclude that: the asymptotic result of Eq. (\ref{61}) holds for all timers $\tau $ simultaneously. Similarly, if the input is DFR then its hazard function is monotone decreasing, hence $\bar{H}\left( \tau \right) > H\left( \infty \right) $ for all timers $\tau $, and thus we conclude that: the asymptotic result of Eq. (\ref{62}) holds for all timers $\tau $ simultaneously.

\section{\label{7} Implications}

We now turn to explore the implications of the asymptotic stochastic-dominance results that were presented in the previous section.

\subsection{\label{hl}The hazard limit}

Evidently, the hazard-function's limit $H\left( \infty \right)$ plays a key role in the asymptotic
results of section \ref{6}. There are three possible values for this hazard limit: zero, positive, and infinite. The `boundary values' straightforwardly yield the following pair of `boundary corollaries'.

\begin{enumerate}
\item[$\bullet $] If $H\left( \infty \right) =\infty $ then the asymptotic result of Eq. (\ref{61}) holds for all timers $\tau $ simultaneously.

\item[$\bullet $] If $H\left(\infty \right) =0$ then the asymptotic result of Eq. (\ref{62}) holds for all timers $\tau $
simultaneously.\footnote{Here we assume that the input's density function is positive-valued over the
positive half-line: $f\left( t\right) >0$ for all $t>0$. In general, the positive hazard-limit case $H\left( \infty \right) =0$ implies that Eq. (\ref{62}) holds for
all timers $\tau >t_{\ast }$ simultaneously, where $t_{\ast }=\inf \{t\geq 0$
$|$ $\bar{F}\left( t\right) <1\}$ is the lower bound of the input's
admissible values.}

\end{enumerate}

The positive hazard-limit case $0<H\left( \infty \right) <\infty $ is more intricate. In this case the asymptotic results of section \ref{6} need not apply simultaneously to all timers $\tau $. Namely (see Fig. 5): for some timer parameters we may have $\bar{H} \left( \tau \right) <H\left( \infty \right) $, yielding Eq. (\ref{61}); and for other timer parameters we may have $\bar{H}\left( \tau \right) >H\left( \infty \right) $, yielding Eq. (\ref{62}). Additional remarks regarding the intricacy of the positive hazard-limit case are detailed in the Methods. 

Last, we note that the limit $H\left( \infty \right)$ can be formulated also in terms of the negative logarithmic derivative of the input's density function: $G\left( t\right) =-f^{\prime }\left( t\right) /f\left( t\right) $. Indeed, assume that the input's density function vanishes at infinity, $\lim_{t\rightarrow \infty}f\left( t\right) =0$. Then, L'Hospital's rule implies that: $H\left( \infty \right) =G\left( \infty \right) $, where $G\left( \infty \right) =\lim_{t\rightarrow \infty }G\left( t\right) $.

\begin{figure}[t]
\centering
\includegraphics[width=9cm]{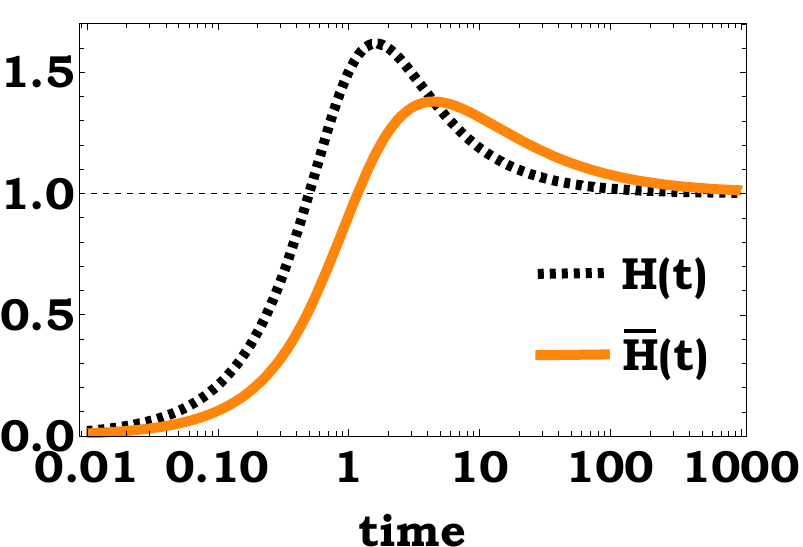}
\caption{An example of the positive hazard-limit case $0<H\left( \infty \right) <\infty $, and an illustration of the optimization results. The example we use here is an input with hazard function $H(t)=(2t+t^2)/(1+t^2)$. We plot this hazard function -- whose limit is $ H\left( \infty \right) =1$ -- in dashed black. Also, we plot the corresponding average function, $\bar{H}\left(t \right) =\frac{1}{t }\int_{0}^{t }H\left( s \right) ds $, in solid orange. With regard to subsection \ref{hl}, note that: the solid orange curve has values that are smaller than the level $ H\left( \infty \right) =1$, as well as values that are larger than this level. With regard to subsection \ref{opti}, note that: the maximum of the solid orange curve is attained at the time point at which this curve intersects the dashed black curve; and that at this time point the dashed black curve is decreasing.}
\label{positive_scenario}
\end{figure}

\subsection{Fast restart}

In this subsection we address `fast restart' -- sharp-restart with sufficiently small timers $\tau \ll1 $. To that end we note that L'Hospital's rule yields the following limit:

\begin{equation}
\lim_{\tau \rightarrow 0}\bar{H}\left( \tau \right) =H\left( 0\right) .  \label{71}
\end{equation}%
As the average $\bar{H}\left( \tau \right) $ is a continuous function of the
timer parameter $\tau $, Eq. (\ref{71}) yields the following pair of `fast-restart corollaries'.

\begin{enumerate}
\item[$\bullet $] If $H\left( 0\right) <H\left( \infty \right) $ then there
exist sufficiently small timers $\tau $ for which the asymptotic result of Eq. (\ref{61}) holds.

\item[$\bullet $] If $H\left( 0\right) >H\left( \infty \right) $ then there
exist sufficiently small timers $\tau $ for which  for which the asymptotic result of Eq. (\ref{62}) holds.
\end{enumerate}

Note that, at zero, the value of the input's hazard function coincides with the value of the input's
density function: $H\left( 0\right) =f\left( 0\right) $. This follows from the fact that, as the input $T$ is positive-valued, the value of its survival function at
zero is one, $\bar{F}\left( 0\right) =1$.

\subsection{Slow restart}

Considering the positive hazard-limit case $0<H\left( \infty \right) <\infty $, in this subsection we address `slow restart' -- sharp restart with sufficiently large timers $\tau \gg1 $. To that end we use the following limit-result:
\begin{equation}
\lim_{\tau \rightarrow \infty }\tau \left[ \bar{H}\left( \tau \right)
-H\left( \infty \right) \right] =\int_{0}^{\infty }\left[ H\left( t\right)
-H\left( \infty \right) \right] dt.  \label{72}
\end{equation}%
The derivation of Eq. (\ref{72}) is detailed in the Methods. As the average $\bar{H}\left( \tau \right) $ is a continuous function of the timer parameter $\tau $, Eq. (\ref{72}) yields the following pair of `slow-restart corollaries'; in these corollaries $I$ denotes the integral appearing on the right-hand side of Eq. (\ref{72}). 

\begin{enumerate}
\item[$\bullet $] If $I<0$ then there exist sufficiently large timers $\tau $
for which the asymptotic result of Eq. (\ref{61}) holds.

\item[$\bullet $] If $I>0$ then there exist sufficiently large timers $\tau $
for which the asymptotic result of Eq. (\ref{62}) holds.
\end{enumerate}

\subsection{Existence} 

Considering the positive hazard-limit case $0<H\left( \infty \right) <\infty $, in this subsection we investigate the very existence of timer parameters that either impede or expedite task-completion. To that end we use the following result:
\begin{equation}
\int_{0}^{\infty }\left[ \bar{H}\left( \tau \right) -H\left( \infty \right) %
\right] f_{1}\left( \tau \right) d\tau =\frac{1}{\mu }-H\left( \infty
\right) ,  \label{73}
\end{equation}%
where $\mu =\mathbf{E}\left[ T\right] $ is the input's mean, and where $%
f_{1}\left( \tau \right) =\frac{1}{\mu }\tau f\left( \tau \right) $. The proof of Eq. (\ref{73}) is detailed in the Methods.\footnote{In the proof we also show that if the limit $H\left( \infty \right) $ is positive then so is the input's mean $\mu $.}  As the term $f_{1}\left( \tau
\right) $ is non-negative valued, Eq. (\ref{73}) yields the following pair
of `existence corollaries'.

\begin{enumerate}
\item[$\bullet $] If $H\left( \infty \right) >\frac{1}{\mu }$ then there
exist timers $\tau $ for which the asymptotic result of Eq. (\ref{61}) holds.

\item[$\bullet $] If $H\left( \infty \right) <\frac{1}{\mu }$ then there
exist timers $\tau $ for which the asymptotic result of Eq. (\ref{62}) holds.
\end{enumerate}

\subsection{\label{opti}Optimization}

Excluding the hazard-limit value $H\left( \infty \right)=\infty $, in this subsection we address the optimization of the right-hand side of Eq. (\ref{60}). Specifically, Eq. (\ref{60}) yields the following pair of optimization observations. If impeding task-completion is a goal, then one would seek to minimize the average $\bar{H}\left( \tau \right) $. And, if expediting task-completion is a goal, then one would seek to maximize the average $\bar{H}\left( \tau \right) $. 

The local minima and the local maxima of the average $\bar{H}\left( \tau \right) $, as a function of the timer
parameter $\tau $, are attained at its critical points: timers $\tau _{c}$ at which the average's derivative vanishes, $\bar{H}^{\prime }\left( \tau
_{c}\right) =0$. A calculation detailed in the Methods implies that the average's derivative is given by $\bar{H}^{\prime }(\tau)=\frac{1}{\tau }\left[ H\left( \tau \right) -\bar{H}\left( \tau \right) \right]$. Consequently, we obtain that the critical points $\tau _{c}$ are the points at which the average $\bar{H}\left( \tau \right) $ intersects the input's hazard function (see Fig. 5):
\begin{equation}
\bar{H}^{\prime }\left( \tau _{c}\right) =0\Leftrightarrow \bar{H} \left( \tau _{c}\right) =H\left( \tau _{c}\right) .
\end{equation}


A calculation detailed in the Methods implies that, at its critical points, the second derivative of the average is given by $\bar{H}^{\prime \prime }(\tau_{c})=\frac{1}{\tau_{c} }H^{\prime }(\tau_{c})$. Thus, for a given critical point $\tau _{c}$, we obtain the following pair optimization conclusions. If the input's hazard function is increasing at the critical point, $H^{\prime}\left( \tau _{c}\right) >0$, then this critical point yields a local minimum of the average $\bar{H}\left( \tau \right) $. Analogously, if the input's hazard function is decreasing at the critical point, $H^{\prime}\left( \tau _{c}\right) <0$, then this critical point yields a local maximum of the average $\bar{H}\left( \tau \right) $ (see Fig. 5).

Last, we note that the local min/max classifications can be formulated also in terms of the negative logarithmic derivative of the input's density function: $G\left( t\right) =-f^{\prime }\left( t\right) /f\left( t\right) $. Indeed, a calculation detailed in the Methods implies that the derivative of the input's hazard function admits the representation ${H}^{\prime }(t)=H(t)[H(t)-G(t)]$. Consequently -- assuming that the input's hazard function is positive at the critical point $\tau _{c}$ -- the local min/max classifications admit the following formulations. Local minimum: if $H(\tau _{c})>G(\tau _{c})$ then the critical point yields a local minimum of the average $\bar{H}\left( \tau \right) $. Local maximum: if $H(\tau _{c})<G(\tau _{c})$ then the critical point yields a local maximum of the average $\bar{H}\left( \tau \right) $.

\section{Summary and manual}

A central issue, in the context of the sharp-restart algorithm, is determining if this algorithm impedes or expedites task-completion. To date, this issue was investigated mainly via the average-behavior perspective: determining if the output's mean is larger than the input's mean, $\mathbf{E}\left[ T_{R}\right] >\mathbf{E}\left[ T\right] $; or if the output's mean is smaller than the input's mean, $\mathbf{E}\left[ T_{R}\right] <\mathbf{E}\left[ T\right] $. Evidently, the average-behavior perspective provides no insight regarding the occurrence likelihood of extremely large completion times. 

Using hazard rates, this paper shifted from the average-behavior perspective to a tail-behavior perspective. Firstly, a compact and neat hazard-rate formulation of the input-to-output map $T\mapsto T_{R}$ was presented, Eq. (\ref{41}). Secondly, using Eq. (\ref{41}), tail-dominance results -- for the
classes of IFR and DFR inputs -- were established. Specifically, if an input is IFR then the output's survival function is larger than that of the input: $\bar{F}_{R}\left( t\right) > \bar{F}\left( t\right) $, over the temporal ray $ t>\tau $. And, if an input is DFR then the output's survival function is smaller than that of the input: $\bar{F}_{R}\left( t\right) < \bar{F}\left( t\right) $, over the temporal ray $ t>\tau $. These tail-dominance results were shown to induce corresponding mean results. 

Thirdly, focusing on the temporal limit $t\rightarrow \infty $, asymptotic tail-dominance results -- for all inputs -- were established. Specifically, universal hazard-rate criteria asserted when the output's survival function decays infinitely slower than the input's survival function: $\lim_{t\rightarrow \infty }\bar{F}_{R}\left(
t\right) /\bar{F}\left( t\right) =\infty$. Also, the universal hazard-rate criteria asserted when the output's survival function decays infinitely faster than the input's survival function: $\lim_{t\rightarrow \infty }\bar{F}_{R}\left( t\right) /\bar{F}\left( t\right) =0 $. The asymptotic tail-dominance results, as well as various corollaries of these universal results, are summarized in Tables I and II.

In the cases of IFR and DFR inputs, as noted above, the effect of sharp restart on tail behavior is in perfect alignment with its effect on average behavior. When stepping out of the realm of inputs with monotone hazard functions the effect of sharp restart on average behavior and on tail behavior can be either aligned or opposite. To determine the joint average-\&-tail effect of sharp restart one has to combine the average-behavior results established in \cite{MPSR1} together with the tail-behavior results summarized in Tables I and II. Doing so yields an average-\&-tail manual that shall be presented now.\\

\begin{center}
{\Large Table I}

\bigskip

\begin{tabular}{||l||l||l||}
\hline\hline
$%
\begin{array}{c}
\ \\ 
\textbf{Timer} \\ 
\
\end{array}%
$ & $%
\begin{array}{c}
\ 
\lim_{t\rightarrow \infty }\frac{\bar{F}_{R}\left( t\right) }{\bar{F}\left(
t\right) }=0 
\
\end{array}%
$ & $%
\begin{array}{c}
\ \\ 
\lim_{t\rightarrow \infty }\frac{\bar{F}_{R}\left( t\right) }{\bar{F}\left(
t\right) }=\infty \\ 
\
\end{array}%
$ \\ \hline\hline
$%
\begin{array}{c}
\ \\ 
Fixed \\ 
\
\end{array}%
$ & $\bar{H}(\tau)>H\left( \infty
\right) $ & $\bar{H}(\tau)<H\left(
\infty \right) $ \\ \hline\hline
$%
\begin{array}{c}
\ \\ 
All \\ 
\
\end{array}%
$ & $H\left( t\right) $ decreasing & $H\left( t\right) $ increasing \\ 
\hline\hline
$%
\begin{array}{c}
\ \\ 
All \\ 
\
\end{array}%
$ & $H\left( \infty \right) =0$ & $H\left( \infty \right) =\infty $ \\ 
\hline\hline
\end{tabular}
\end{center}

\textbf{Table I}: Summary of key asymptotic results. The table specifies, in terms of the input's hazard function $H(t)$, universal criteria leading to the limits $\lim_{t\rightarrow \infty }\bar{F}_{R}\left(t\right) /\bar{F}\left( t\right) =0$ and $\lim_{t\rightarrow \infty }\bar{F}_{R}\left( t\right) /\bar{F}\left( t\right) =\infty $. The criteria appearing in the first row apply to any fixed (specific) timer $\tau $; in this row $\bar{H}(\tau)=\frac{1}{\tau }\int_{0}^{\tau }H\left( t\right) dt$. The criteria appearing in the second and third rows apply to all timers, $0< \tau < \infty$, simultaneously. See section \ref{6} for the details of these universal hazard-rate criteria.\\

\newpage

\begin{center}
{\Large Table II}

\bigskip

\begin{tabular}{||l||l||l||}
\hline\hline
$%
\begin{array}{c}
\ \\ 
\textbf{Timer} \\ 
\
\end{array}%
$ & $%
\begin{array}{c}
\ \\ 
\lim_{t\rightarrow \infty }\frac{\bar{F}_{R}\left( t\right) }{\bar{F}\left(
t\right) }=0 \\ 
\
\end{array}%
$ & $%
\begin{array}{c}
\ \\ 
\lim_{t\rightarrow \infty }\frac{\bar{F}_{R}\left( t\right) }{\bar{F}\left(
t\right) }=\infty \\ 
\
\end{array}%
$ \\ \hline\hline
$%
\begin{array}{c}
\ \\ 
General \\ 
\
\end{array}%
$ & $ \frac{1}{\mu }> H\left( \infty \right)$ & $ \frac{1}{\mu } < H\left( \infty \right) $ \\ \hline\hline
$%
\begin{array}{c}
\ \\ 
Small \\ 
\
\end{array}%
$ & $H\left( 0\right) >H\left( \infty \right) $ & $H\left( 0\right) <H\left(
\infty \right) $ \\ \hline\hline
$%
\begin{array}{c}
\ \\ 
Large \\ 
\
\end{array}%
$ & $I>0$ & $I<0$ \\ \hline\hline
\end{tabular}
\end{center}

\textbf{Table II}: Summary of key existence results for the positive hazard-limit case $0<H\left( \infty \right) <\infty $. The table specifies, in terms of the input's hazard function $H(t)$, universal criteria that determine the very existence of timers $\tau $ for which the limits $\lim_{t\rightarrow \infty }\bar{F}_{R}\left(t\right) /\bar{F}\left( t\right) =0$ and $\lim_{t\rightarrow \infty }\bar{F}_{R}\left( t\right) /\bar{F}\left( t\right) =\infty $ hold. First row: general timers, $0< \tau < \infty $; in this row $\mu=\mathbf{E}\left[ T\right]$. Second row: sufficiently small timers, $\tau \ll 1 $. Third row: sufficiently large timers, $\tau \gg 1 $; in this row $I=\int_{0}^{\infty }\left[ H\left( t\right) -H\left( \infty \right) %
\right] dt$. See section \ref{7} for the details of these universal hazard-rate criteria.

\bigskip

The manual provides a universal classification of sharp restart from a joint average-behavior and tail-behavior perspectives. The manual shows how essential it is -- when applying sharp restart -- to jointly use universal average-behavior and tail-behavior criteria. In particular, the manual underscores the indispensability of the novel tail-behavior results established here. The manual addresses fast restart, slow restart, and mean-critical restart.

Throughout the manual the following jargon is used. Regarding average behavior: sharp restart with a given timer is \emph{mean-detrimental} if $\mathbf{E}\left[ T_{R}\right] > \mathbf{E}\left[ T\right] $, and is \emph{mean-beneficial} if $\mathbf{E}\left[ T_{R}\right] < \mathbf{E}\left[ T\right] $. Analogously, regarding tail behavior: sharp restart with a given timer is \emph{tail-detrimental} if $\lim_{t\rightarrow \infty }\bar{F}_{R}\left(t\right) /\bar{F}\left( t\right) =\infty$, and is \emph{tail-beneficial} if $\lim_{t\rightarrow \infty }\bar{F}_{R}\left(t\right) /\bar{F}\left( t\right) =0$. Also, as above, the shorthand notation $\mu=\mathbf{E}\left[ T\right]$ is used.

\subsection{\label{91} Average-\&-tail manual for fast restart}

Consider `fast restart' -- sharp restart with sufficiently small timers. Fast-restart criteria established in \cite{MPSR1}, together with fast-restart criteria established here, yield four possible scenarios regarding the interplay between average behavior and tail behavior; these four scenarios are detailed in Table III. The scenarios in which fast restart has an opposite effect on average behavior and on tail behavior are exemplified below.\\
\newpage

\begin{center}
{\Large Table III}

\bigskip 

\begin{tabular}{||c||c||c||}
\hline\hline
& $%
\begin{array}{c}
\\ 
\textbf{Tail B} 
\\ 
\end{array}%
$ & $%
\begin{array}{c}
\\ 
\textbf{Tail D} 
\\ 
\end{array}%
$ \\ \hline\hline
$%
\begin{array}{c}
\\ 
\textbf{Mean B} 
\\ 
\end{array}%
$ & $%
\begin{array}{c}
\\ 
H\left( \infty \right) <H(0) \\ 
\\ 
\frac{1}{\mu }<H(0) 
\\ 
\end{array}%
$ & $\frac{1}{\mu }<H(0)<H\left( \infty \right) $ \\ \hline\hline
$%
\begin{array}{c}
\\ 
\textbf{Mean D} 
\\ 
\end{array}%
$ & $H\left( \infty \right) <H(0)<\frac{1}{\mu }$ & $%
\begin{array}{c}
\\ 
H(0)<H\left( \infty \right)  \\ 
\\ 
H(0)<\frac{1}{\mu } 
\\ 
\end{array}%
$ \\ \hline\hline
\end{tabular}
\end{center}

\textbf{Table III:} The joint effect of fast restart -- sharp restart with sufficiently small timers $\tau \ll 1 $ -- on average behavior and on tail behavior. The rows specify when fast restart is mean-beneficial (Mean B), and when it is mean-detrimental (Mean D). The columns specify when fast restart is tail-beneficial (Tail B), and when it is tail-detrimental (Tail D).

\subsection{\label{92} Average-\&-tail manual for slow restart}

Consider `slow restart' -- sharp restart with sufficiently large timers. Slow-restart criteria established in \cite{MPSR1}, together with slow-restart criteria established here, imply that the following cases lead to markedly different outcomes: exploding hazard-limit, $H(\infty)=\infty$; vanishing hazard-limit, $H(\infty)=0$; and positive hazard-limit, $0<H(\infty)<\infty$. In the exploding and vanishing cases the effect of slow restart on tail behavior is aligned with its effect on average behavior: if $H(\infty)=\infty$ then slow restart is both mean-detrimental and tail-detrimental; and, if $H(\infty)=0$ then slow restart is both mean-beneficial and tail-beneficial. Intricacy appears in the positive case: if $0<H(\infty)<\infty$ then there are four possible scenarios regarding the interplay between average behavior and tail behavior; these four scenarios are detailed in Table IV.

Examples of inputs with an exploding hazard-limit $H(\infty)=\infty$ include: the Gompertz input of Fig. 3, and the  Rayleigh input (over the temporal ray $t \geq l$) in Example A below. Examples of inputs with a vanishing hazard-limit $H(\infty)=0$ include \cite{MPSR1}: Lognormal completion times, which are observed as service times in call centers \cite{BGM}-\cite{GT2}; Stretched-Exponential relaxation times, which are prevalent in anomalous relaxation \cite{WW}-\cite{CK1}; and first-passage times with asymptotic power-law tails, which arise from Brownian motion and from Langevin dynamics on logarithmic potentials \cite{Diff9,AJBray}. With regard to the positive hazard-limit case $0<H(\infty)<\infty$, the scenarios in which slow restart has an opposite effect on average behavior and on tail behavior are exemplified below.\\
\newpage

\begin{center} 
{\Large Table IV}

\bigskip

\begin{tabular}{||c||c||c||}
\hline\hline
& $%
\begin{array}{c}
\\ 
\textbf{Tail B} 
\\ 
\end{array}%
$ & $%
\begin{array}{c}
\\ 
\textbf{Tail D} 
\\ 
\end{array}%
$ \\ \hline\hline
$%
\begin{array}{c}
\\ 
\textbf{Mean B} 
\\ 
\end{array}%
$ & $%
\begin{array}{c}
\\ 
H\left( \infty \right) <\frac{1}{\mu } \\ 
\\ 
I>0 
\\ 
\end{array}%
$ & $%
\begin{array}{c}
\\ 
H\left( \infty \right) <\frac{1}{\mu } \\ 
\\ 
I<0 
\\ 
\end{array}%
$ \\ \hline\hline
$%
\begin{array}{c}
\\ 
\textbf{Mean D} 
\\ 
\end{array}%
$ & $%
\begin{array}{c}
\\ 
H\left( \infty \right) >\frac{1}{\mu } \\ 
\\ 
I>0 
\\ 
\end{array}%
$ & $%
\begin{array}{c}
\\ 
H\left( \infty \right) >\frac{1}{\mu } \\ 
\\ 
I<0 \\ 
\end{array}%
$ \\ \hline\hline
\end{tabular}
\end{center}

\textbf{Table IV}: The joint effect of slow restart -- sharp restart with sufficiently large timers $\tau \gg 1 $ -- on average behavior and on tail behavior, in the positive hazard-limit case $0<H(\infty)<\infty$. The rows specify when slow restart is mean-beneficial (Mean B), and when it is mean-detrimental (Mean D). The columns specify when slow restart is tail-beneficial (Tail B), and when it is tail-detrimental (Tail D). As above, $I=\int_{0}^{\infty }\left[ H\left( t\right) -H\left( \infty \right) 
\right] dt$.

\subsection{\label{93} Average-\&-tail manual for mean-critical restart}

Address the output's mean as a function of the timer parameter, $\mathbf{E}\left[ T_{R}\right]=M(\tau)$ ($0<\tau<\infty$). Evidently, the mean function $M(\tau)$ can attain its global minimum either at the boundaries of the positive half-line (i.e. in the limits $\tau \to 0$ and $\tau \to \infty$), or at local minima (if such exist). In turn, local minima can occur only at `mean-critical' timers: points $\tau_{c}$ at which the derivative of the mean function vanishes, $M'(\tau_{c})=0$. For mean-critical timers: criteria established in \cite{MPSR1}, together with criteria established here, yield four possible scenarios regarding the interplay between average behavior and tail behavior; these four scenarios are detailed in Table V.

\subsection{\label{94} Examples}

With regard to the manual described above, the following examples demonstrate scenarios in which sharp restart has an opposite effect on average behavior and on tail behavior: mean-beneficial and tail-detrimental; or, conversely, mean-detrimental and tail-beneficial. The examples share a common underlying task-completion scheme that is described as follows. A task arrives at a service station and, upon arrival, it is attended by a primary server. If the primary server does not manage to complete the task within $l$ time units, the task is transferred to a secondary server. Upon transfer, the secondary server starts attending the task afresh.

\clearpage

\begin{center}
{\Large Table V}

\bigskip 

\begin{tabular}{||c||c||c||}
\hline\hline
& $%
\begin{array}{c}
\\ 
\textbf{Tail B} 
\\ 
\end{array}%
$ & $%
\begin{array}{c}
\\ 
\textbf{Tail D} 
\\ 
\end{array}%
$ \\ \hline\hline
$%
\begin{array}{c}
\\ 
\textbf{Mean B} 
\\ 
\end{array}%
$ & $%
\begin{array}{c}
\\ 
H\left( \tau _{c}\right) >\frac{1}{\mu } \\ 
\\ 
\bar{H}\left( \tau _{c}\right) >H\left( \infty \right)  \\ 
\end{array}%
$ & $%
\begin{array}{c}
\\ 
H\left( \tau _{c}\right) >\frac{1}{\mu } \\ 
\\ 
\bar{H}\left( \tau _{c}\right) <H\left( \infty \right)  \\ 
\end{array}%
$ \\ \hline\hline
$%
\begin{array}{c}
\\ 
\textbf{Mean D}
\\ 
\end{array}%
$ & $%
\begin{array}{c}
\\ 
H\left( \tau _{c}\right) <\frac{1}{\mu } \\ 
\\ 
\bar{H}\left( \tau _{c}\right) >H\left( \infty \right)  
\\ 
\end{array}%
$ & $%
\begin{array}{c}
\\ 
H\left( \tau _{c}\right) <\frac{1}{\mu } \\ 
\\ 
\bar{H}\left( \tau _{c}\right) <H\left( \infty \right)  \\ 
\end{array}%
$ \\ \hline\hline
\end{tabular}
\end{center}

\textbf{Table V}: The joint effect of sharp restart with mean-critical timers -- $\tau_{c}$ such that $M'(\tau_{c})=0$ -- on average behavior and on tail behavior. The rows specify when sharp restart is mean-beneficial (Mean B), and when it is mean-detrimental (Mean D). The columns specify when sharp restart is tail-beneficial (Tail B), and when it is tail-detrimental (Tail D).\\

\emph{Example A}: the scenario $ \frac{1}{\mu} < H(0) < H(\infty)$ of Table III, in which fast restart is mean-beneficial and tail-detrimental. This scenario is demonstrated via an input whose statistics are Exponential up to the positive level $l$, and Rayleigh above this level. Specifically, the input's hazard function is (see Fig. 6A): $H(t)=r$ over the temporal interval $0<t<l$, and $H(t)=s(t-l)$ over the temporal ray $t\geq l$; the Exponential rate $r$ and the Rayleigh slope $s$ are positive parameters. A calculation of the corresponding mean $\mu$ implies that the scenario under consideration is met if and only if $\pi r^2 > 2s$.

\emph{Example B}: the scenario $H(\infty) < H(0) < \frac{1}{\mu}$ of Table III, in which fast restart is mean-detrimental and tail-beneficial. This scenario is demonstrated via an input whose statistics are Exponential up to the positive level $l$, and type-I Pareto above this level. Specifically, the input's hazard function is (see Fig. 6B): $H(t)=r$ over the temporal interval $0<t<l$, and $H(t)=p/t$ over the temporal ray $t\geq l$; the Exponential rate $r$ is a positive parameter, and the Pareto power $p$ is in the finite-mean range $p>1$. A calculation of the corresponding mean $\mu$ implies that the scenario under consideration is met if and only if $p-1>rl$.


\begin{figure}[t]
\centering
\includegraphics[width=12cm]{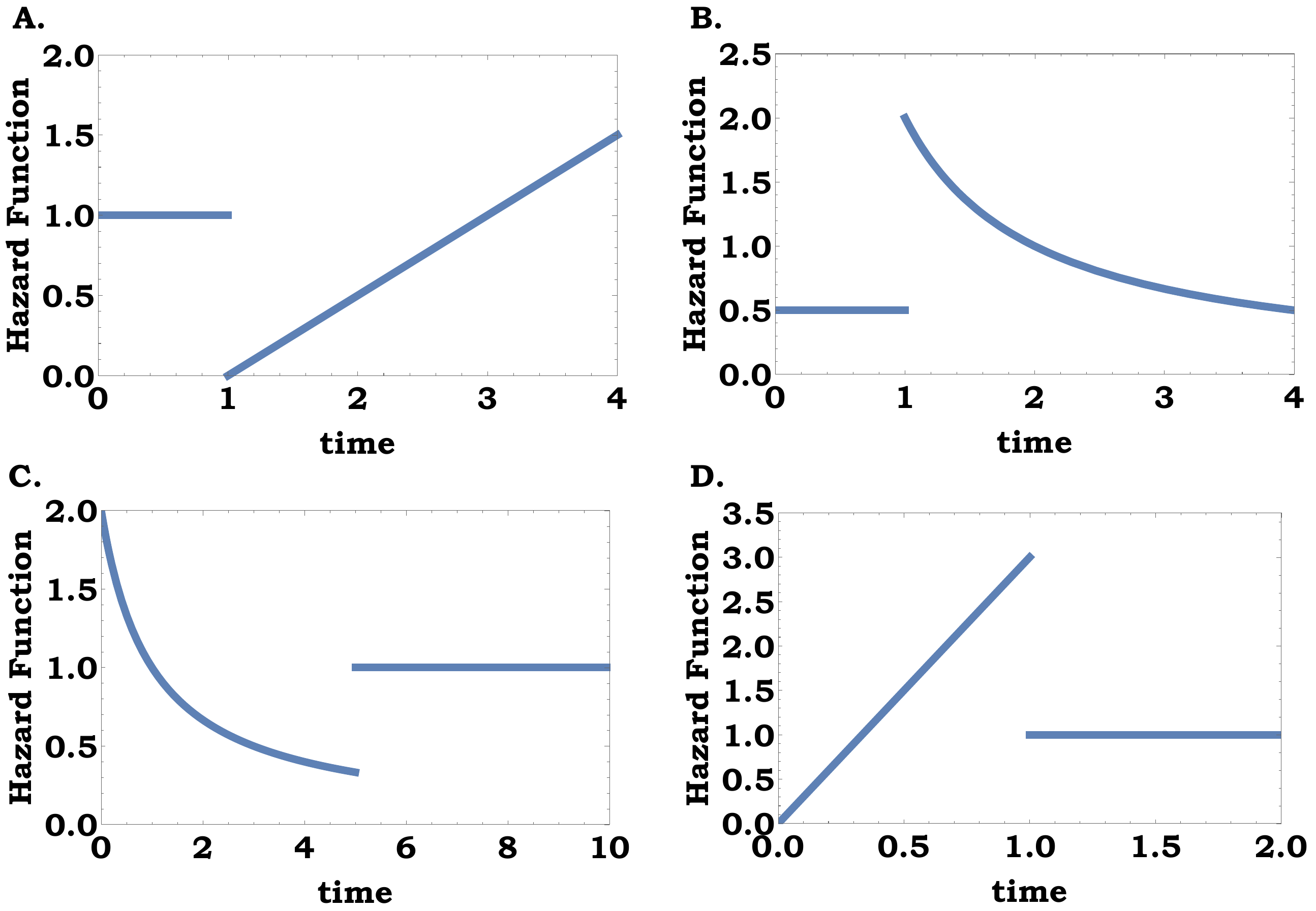}
\caption{Illustration of examples A-D of subsection \ref{94}. The figures depict the hazard functions of inputs that exhibit one type of statistics up to a positive level $l$, and another type of statistics above the level. Panel A: Exponential statistics with rate $r=1$ up to the level $l=1$, and Rayleigh statistics with slope $s=1/2$ above the level. Panel B: Exponential statistics with rate $r=1/2$ up to the level $l=1$, and type-I Pareto statistics with power $p=2$ above the level. Panel C: type-II Pareto statistics with power $p=2$ up to the level $l=5$, and Exponential statistics with rate $r=1$ above the level. Panel D: Rayleigh statistics with slope $s=3$ up to the level $l=1$, and Exponential statistics with rate $r=1$ above the level.}
\label{Hazard_fig}
\end{figure}

\emph{Example C}: the scenario $ H(\infty) < \frac{1}{\mu}$ \& $I<0$ of Table IV, in which slow restart is mean-beneficial and tail-detrimental. This scenario is demonstrated via an input whose statistics are type-II Pareto up to the positive level $l$, and Exponential above this level. Specifically, the input's hazard function is (see Fig. 6C): $H(t)=p/(1+t)$ over the temporal interval $0<t<l$, and $H(t)=r$ over the temporal ray $t\geq l$; the Exponential rate $r$ is a positive parameter, and the Pareto power is $p=2$. A calculation of the corresponding mean $\mu$ and integral $I$ implies that the scenario under consideration is met if and only if $2\ln{(1+l)}/l < r < 1+1/(1+l)$. 

\emph{Example D}: the scenario $ H(\infty) > \frac{1}{\mu}$ \& $I>0$ of Table IV, in which slow restart is mean-detrimental and tail-beneficial. This scenario is demonstrated via an input whose statistics are Rayleigh up to the positive level $l$, and Exponential above this level. Specifically, the input's hazard function is (see Fig. 6D): $H(t)=st$ over the temporal interval $0<t<l$, and $H(t)=r$ over the temporal ray $t\geq l$; the  Rayleigh slope $s$ and the Exponential rate $r$ are positive parameters. A calculation of the corresponding mean $\mu$ and integral $I$ implies that -- when $sl^2$ is sufficiently large -- the scenario under consideration is met if and only if $\sqrt{s/\pi} < r < sl/2 $.

Examples A-D display two hazard-function features: non-monotonicity and discontinuity. As established above, if the input's hazard function is monotone then the effect of sharp restart on average behavior and on tail behavior is aligned. Hence, non-monotonicity is an essential feature in examples A-D. On the other hand, the discontinuity feature is merely technical: its only purpose is to ease the examples' calculations. Indeed, examples A-D can be tinkered to have smooth hazard functions, while maintaining their opposite average-\&-tail behaviors.\\

\textbf{Acknowledgments}. The authors thanks Ofek Lauber Bonomo for help with preprations of figures in this paper. Shlomi Reuveni acknowledges support from the Azrieli Foundation, from the Raymond and Beverly Sackler Center for Computational Molecular and Materials Science at Tel Aviv University, and from the Israel Science Foundation (grant No. 394/19).

\section{Methods}

\subsection{Proof of Eq. (\protect\ref{60})}

Consider the ratio 
\begin{equation}
\rho \left( t\right) =\frac{\bar{F}_{R}\left( t\right) }{\bar{F}\left(
t\right) }  \label{A200}
\end{equation}%
at the time $t=n\tau +u$ where: $n$ is a non-negative integer, and $u$ is a
fixed value in the range $0\leq u<\tau $. As the input's and output's
survival functions are monotone decreasing we have%
\begin{equation}
\bar{F}\left[ \left( n+1\right) \tau \right] \leq \bar{F}\left( n\tau
+u\right) \leq \bar{F}\left( n\tau \right) ,  \label{A201}
\end{equation}%
and 
\begin{equation}
\bar{F}_{R}\left[ \left( n+1\right) \tau \right] \leq \bar{F}_{R}\left(
n\tau +u\right) \leq \bar{F}_{R}\left( n\tau \right) .  \label{A202}
\end{equation}%
In turn, Eqs. (\ref{A201})-(\ref{A202}) imply that%
\begin{equation}
\frac{\bar{F}_{R}\left[ \left( n+1\right) \tau \right] }{\bar{F}\left( n\tau
\right) }\leq \rho \left( n\tau +u\right) \leq \frac{\bar{F}_{R}\left( n\tau
\right) }{\bar{F}\left[ \left( n+1\right) \tau \right] } .
\label{A203}
\end{equation}

In terms of the input's hazard function, the input's survival function is
given by $\bar{F}\left( t\right) =\exp \{-\int_{0}^{t}H(s)ds\}$. Hence, using Eq. (\ref{31}): for the lower bound appearing on the left-hand side of \eref{A203} we have
\begin{equation}
\frac{\bar{F}_{R}\left[ \left( n+1\right) \tau \right] }{\bar{F}\left( n\tau
\right) }=\frac{\bar{F}\left( \tau \right) ^{n+1}}{\bar{F}\left( n\tau
\right) }=\frac{\exp \left[ -\left( n+1\right) \int_{0}^{\tau }H\left(
s\right) ds\right] }{\exp \left[ -\int_{0}^{n\tau }H\left( s\right) ds\right]
} ,  \label{A211}
\end{equation}%
and for the upper bound appearing on the right-hand side of \eref{A203} we have 
\begin{equation}
\frac{\bar{F}_{R}\left( n\tau \right) }{\bar{F}\left[ \left( n+1\right) \tau %
\right] }=\frac{\bar{F}\left( \tau \right) ^{n}}{\bar{F}\left[ \left(
n+1\right) \tau \right] }=\frac{\exp \left[ -n\int_{0}^{\tau }H\left(
s\right) ds\right] }{\exp \left[ -\int_{0}^{\left( n+1\right) \tau }H\left(
s\right) ds\right] }.  \label{A212}
\end{equation}%
In turn, Eq. (\ref{A203}) and Eqs. (\ref{A211})-(\ref{A212}) imply that%
\begin{equation}
\left. 
\begin{array}{l}
\int_{0}^{n\tau }H\left( s\right) ds-\left( n+1\right) \int_{0}^{\tau
}H\left( s\right) ds \\ 
\ \\ 
\leq \ln \left[ \rho \left( n\tau +u\right) \right] \\ 
\ \\ 
\leq \int_{0}^{\left( n+1\right) \tau }H\left( s\right) ds-n\int_{0}^{\tau
}H\left( s\right) ds.
\end{array}%
\right.  \label{A213}
\end{equation}

Introduce the average 
\begin{equation}
\bar{H}\left( l\right) =\frac{1}{l}\int_{0}^{l}H\left( s\right) ds
\label{A221}
\end{equation}%
($l>0$). In terms of this average we can re-write the left-hand side and the
right-hand side of Eq. (\ref{A213}) as follows:%
\begin{equation}
\left. 
\begin{array}{l}
\int_{0}^{n\tau }H\left( s\right) ds-\left( n+1\right) \int_{0}^{\tau
}H\left( s\right) ds \\ 
\ \\ 
=n\tau \left[ \frac{1}{n\tau }\int_{0}^{n\tau }H\left( s\right) ds-\frac{n+1%
}{n}\frac{1}{\tau }\int_{0}^{\tau }H\left( s\right) ds\right] \\ 
\ \\ 
=n\tau \left[ \bar{H}\left( n\tau \right) -\frac{n+1}{n}\bar{H}\left( \tau
\right) \right] ,
\end{array}%
\right.  \label{A222}
\end{equation}%
and%
\begin{equation}
\left. 
\begin{array}{l}
\int_{0}^{\left( n+1\right) \tau }H\left( s\right) ds-n\int_{0}^{\tau
}H\left( s\right) ds \\ 
\ \\ 
=\left( n+1\right) \tau \left[ \frac{1}{\left( n+1\right) \tau }%
\int_{0}^{\left( n+1\right) \tau }H\left( s\right) ds-\frac{n}{n+1}\frac{1}{%
\tau }\int_{0}^{\tau }H\left( s\right) ds\right] \\ 
\ \\ 
=\left( n+1\right) \tau \left\{ \bar{H}\left[ (n+1)\tau \right] -\frac{n}{n+1%
}\bar{H}\left( \tau \right) \right\} .%
\end{array}%
\right.  \label{A223}
\end{equation}

L'Hospital's rule implies that%
\begin{equation}
\lim_{l\rightarrow \infty }\bar{H}\left( l\right) =\frac{1}{l}%
\int_{0}^{l}H\left( s\right) ds=\lim_{l\rightarrow \infty }H\left( l\right)
=H\left( \infty \right) .  \label{A231}
\end{equation}%
Eq. (\ref{A231}) implies that%
\begin{equation}
\lim_{n\rightarrow \infty }\left[ \bar{H}\left( n\tau \right) -\frac{n+1}{n}%
\bar{H}\left( \tau \right) \right] =H\left( \infty \right) -\bar{H}\left(
\tau \right) ,  \label{A232}
\end{equation}%
and that%
\begin{equation}
\lim_{n\rightarrow \infty }\left\{ \bar{H}\left[ (n+1)\tau \right] -\frac{n}{%
n+1}\bar{H}\left( \tau \right) \right\} =H\left( \infty \right) -\bar{H}%
\left( \tau \right) .  \label{A233}
\end{equation}

Eq. (\ref{A213}) and Eqs. (\ref{A222})-(\ref{A223}) imply that 
\begin{equation}
\left. 
\begin{array}{l}
\frac{n\tau }{n\tau +u}\left[ \bar{H}\left( n\tau \right) -\frac{n+1}{n}\bar{%
H}\left( \tau \right) \right] \\ 
\ \\ 
\leq \frac{\ln \left[ \rho \left( n\tau +u\right) \right] }{n\tau +u} \\ 
\ \\ 
\leq \frac{\left( n+1\right) \tau }{n\tau +u}\left\{ \bar{H}\left[ (n+1)\tau %
\right] -\frac{n}{n+1}\bar{H}\left( \tau \right) \right\} .
\end{array}%
\right.  \label{A234}
\end{equation}%
Taking the limit $n\rightarrow \infty $ in Eq. (\ref{A234}), while using the
limits of Eqs. (\ref{A232})-(\ref{A233}), yields%
\begin{equation}
H\left( \infty \right) -\bar{H}\left( \tau \right) \leq \lim_{n\rightarrow
\infty }\frac{\ln \left[ \rho \left( n\tau +u\right) \right] }{n\tau +u}\leq
H\left( \infty \right) -\bar{H}\left( \tau \right) .  \label{A235}
\end{equation}%
As Eq. (\ref{A235}) holds for any fixed value $u$ (in the range $0\leq
u<\tau $) it proves Eq. (\ref{60}):%
\begin{equation}
\lim_{t\rightarrow \infty }\frac{1}{t}\ln \left[ \rho \left( t\right) \right]
=H\left( \infty \right) -\bar{H}\left( \tau \right) .  \label{A236}
\end{equation}

\subsection{Proofs of Eqs. (\protect\ref{61}) and (\protect\ref{62})}

The proof of Eq. (\protect\ref{60}) yielded Eq. (\ref{A235}). As we shall now argue, Eq. (\ref{A235}) leads to Eqs. (\protect\ref{61}) and (\protect\ref{62}).

If $\bar{H}\left( \tau \right) <H\left( \infty \right) $ then Eq. (\ref{A235}%
) implies that%
\begin{equation}
\infty \leq \lim_{n\rightarrow \infty }\ln \left[ \rho \left( n\tau
+u\right) \right] \leq \infty ,  \label{A243}
\end{equation}%
and hence 
\begin{equation}
\lim_{n\rightarrow \infty }\rho \left( n\tau +u\right) =\infty .
\label{A244}
\end{equation}%
As Eq. (\ref{A244}) holds for any fixed value $u$ (in the range $0\leq
u<\tau $) it proves Eq. (\ref{61}):%
\begin{equation}
\lim_{t\rightarrow \infty }\rho \left( t\right) =\infty .
\label{A245}
\end{equation}

If $\bar{H}\left( \tau \right) >H\left( \infty \right) $ then Eq. (\ref{A235}%
) implies that%
\begin{equation}
-\infty \leq \lim_{n\rightarrow \infty }\ln \left[ \rho \left( n\tau
+u\right) \right] \leq -\infty ,  \label{A253}
\end{equation}%
and hence 
\begin{equation}
\lim_{n\rightarrow \infty }\rho \left( n\tau +u\right) =0 .
\label{A254}
\end{equation}%
As Eq. (\ref{A254}) holds for any fixed value $u$ (in the range $0\leq
u<\tau $) it proves Eq. (\ref{62}):%
\begin{equation}
\lim_{t\rightarrow \infty }\rho \left( t\right) =0.  \label{A255}
\end{equation}

\subsection{The positive hazard-limit case}

Consider the positive hazard-limit $0<H\left( \infty \right) <\infty $. In this case one may intuitively assume that the input's survival function is
asymptotically exponential:
\begin{equation}
\bar{F}\left( t\right) \approx \exp [-H\left(\infty \right) t] , \label{76}
\end{equation}%
where the asymptotic equivalence is in the limit $t\rightarrow \infty $. However -- as we shall now argue -- Eq. (\ref{76}) does \emph{not} hold in general.

Using the representation of the input's survival function in terms of its hazard function, $\bar{F}\left( t\right) =\exp \{-\int_{0}^{t}H(s)ds\}$, we have:
\begin{equation}
\left. 
\begin{array}{l}
\lim_{t\rightarrow \infty }\left\{ \exp \left[ H\left( \infty \right) t%
\right] \cdot \bar{F}\left( t\right) \right\}  \\ 
\ \\ 
=\lim_{t\rightarrow \infty }\left\{ \exp \left[ \int_{0}^{t}H\left( \infty
\right) ds\right] \cdot \exp \left[ -\int_{0}^{t}H(s)ds\right] \right\}  \\ 
\ \\ 
=\lim_{t\rightarrow \infty }\exp \left\{ -\int_{0}^{t}\left[ H\left(
s\right) -H\left( \infty \right) \right] ds\right\}  \\ 
\ \\ 
=\exp \left\{ -\lim_{t\rightarrow \infty }\int_{0}^{t}\left[ H\left(
s\right) -H\left( \infty \right) \right] ds\right\}  \\ 
\ \\ 
=\exp \left\{ -\int_{0}^{\infty }\left[ H\left( s\right) -H\left( \infty
\right) \right] ds\right\} .%
\end{array}%
\right.   \label{A270}
\end{equation}%
Consequently, denoting by $I=\int_{0}^{\infty }\left[ H\left( s\right) -H\left( \infty
\right) \right] ds$ the integral appearing on the bottom line of Eq. (\ref{A270}), we assert that: Eq. (\ref{76}) holds if and only if the integral $I$ is convergent, $-\infty <I<\infty $.

As an illustrative example consider an input $T$ whose statistical distribution is governed by the survival function $\bar{F}\left(
t\right) =\left( 1+t\right) ^{1-p}\exp (-t)$ ($t\geq 0$), where $p$ is a
positive parameter. In turn, the input's hazard function is $H\left(
t\right) =(p+t)/(1+t)$ ($t\geq 0$), and hence $H\left( \infty \right) =1$. For all parameter
values $p\neq 1$ the input's survival function is \emph{not} asymptotically
exponential, and indeed: $I=-\infty $ when $p<1$, and $I=\infty $ when $p>1$. On the other hand, for $p=1$ the input's survival function is exponential, and we have $I=0$.

\subsection{Proofs of Eqs. (\ref{72}) and (\ref{73})}

Considering the positive hazard-limit case $0<H\left( \infty \right) <\infty $, note that%
\begin{equation}
\left. 
\begin{array}{l}
\tau \left[ \bar{H}\left( \tau \right) -H\left( \infty \right) \right] =\tau %
\left[ \frac{1}{\tau }\int_{0}^{\tau }H\left( s\right) ds\right] -\tau
H\left( \infty \right)  \\ 
\ \\ 
=\int_{0}^{\tau }H\left( s\right) ds-\int_{0}^{\tau }H\left( \infty \right)
ds=\int_{0}^{\tau }\left[ H\left( s\right) -H\left( \infty \right) \right] ds.
\end{array}%
\right.   \label{A271}
\end{equation}%
In turn, taking the limit $\tau \rightarrow \infty $ in Eq. (\ref{A271})
yields Eq. (\ref{72}):%
\begin{equation}
\lim_{\tau \rightarrow \infty }\tau \left[ \bar{H}\left( \tau \right)
-H\left( \infty \right) \right] =\int_{0}^{\infty }\left[ H\left( s\right)
-H\left( \infty \right) \right] ds.  \label{A272}
\end{equation}

We now move to the proof of Eq. (\ref{73}). Set $K\left( t\right) $ to be the running integral of the input's hazard
function, i.e.%
\begin{equation}
K\left( t\right) =\int_{0}^{t}H\left( s\right) ds  \label{A273}
\end{equation}%
($t\geq 0$). As the input's survival function $\bar{F}\left( t\right) $
decreases monotonically from $\bar{F}\left( 0\right) =1$ to $%
\lim_{t\rightarrow \infty }\bar{F}\left( t\right) =0$, and as $\bar{F}\left(
t\right) =\exp \left[ -K\left( t\right) \right] $, we obtain that: the
function $K\left( t\right) $ increases monotonically from $K\left( 0\right)
=0$ to $\lim_{t\rightarrow \infty }K\left( t\right) =\infty $. Assume that
there exists a positive level $l_{\ast }$ above which the input's density
function is positive-valued: $f\left( t\right) >0$ for all $t>l_{\ast }$.
Note that, over the ray $\left( l_{\ast },\infty \right) $, the function 
$K\left( t\right) $ has an inverse function $K^{-1}\left( \cdot \right) $.

Set an arbitrary level $l>l_{\ast }$. In terms of the input's survival
function the input's mean $\mu =\mathbf{E}\left[ T\right] $ admits the
representation%
\begin{equation}
\mu =\int_{0}^{\infty }\bar{F}\left( t\right) dt=\int_{0}^{l}\bar{F}\left(
t\right) dt+\int_{l}^{\infty }\bar{F}\left( t\right) dt.
\label{A274}
\end{equation}%
Using the fact that $\bar{F}\left( t\right) =\exp \left[ -K\left( t\right) %
\right] $, and the change-of-variables $u=K\left( t\right) $, we have%
\begin{equation}
\left. 
\begin{array}{l}
\int_{l}^{\infty }\bar{F}\left( t\right) dt=\int_{l}^{\infty }\exp \left[
-K\left( t\right) \right] dt \\ 
\ \\ 
=\int_{K(l)}^{\infty }\exp \left( -u\right) \frac{1}{H\left[ K^{-1}\left(
u\right) \right] }du.
\end{array}%
\right.   \label{A275}
\end{equation}%
Also, note that 
\begin{equation}
\lim_{u\rightarrow \infty }H\left[ K^{-1}\left( u\right) \right]
=\lim_{t\rightarrow \infty }H\left( t\right) =H\left( \infty \right).  \label{A276}
\end{equation}%
If $0<H\left( \infty \right) <\infty $ then Eq. (\ref{A276}) implies that
the integral appearing on the right-hand side of Eq. (\ref{A275}) is
convergent. Consequently, we obtain the following implication: $0<H\left( \infty \right) <\infty \Rightarrow 0<\mu <\infty$. 

Considering the positive hazard-limit case $0<H\left( \infty \right) <\infty $, introduce the
function 
\begin{equation}
f_{1}\left( t\right) =\frac{1}{\mu }tf\left( t\right)   \label{A260}
\end{equation}%
($t\geq 0$). Note that $f_{1}\left( t\right) $ is a density function: it is
non-negative, $f_{1}\left( t\right) \geq 0$; and it is normalized, $%
\int_{0}^{\infty }f_{1}\left( t\right) dt=1$. In turn, note that 
\begin{equation}
\left. 
\begin{array}{l}
\int_{0}^{\infty }\left[ \bar{H}\left( \tau \right) -H\left( \infty \right) %
\right] f_{1}\left( \tau \right) d\tau  \\ 
\ \\ 
=\int_{0}^{\infty }\bar{H}\left( \tau \right) f_{1}\left( \tau \right) d\tau
-\int_{0}^{\infty }H\left( \infty \right) f_{1}\left( \tau \right) d\tau  \\ 
\ \\ 
=\frac{1}{\mu }\int_{0}^{\infty }\bar{H}\left( \tau \right) \left[ \tau
f\left( \tau \right) \right] d\tau -H\left( \infty \right) .
\end{array}%
\right.   \label{A261}
\end{equation}%
Using the definition of the average $\bar{H}\left( \tau \right) $, as well
as the definitions of the input's hazard function and survival function, we
have%
\begin{equation}
\left. 
\begin{array}{l}
\int_{0}^{\infty }\bar{H}\left( \tau \right) \left[ \tau f\left( \tau
\right) \right] d\tau =\int_{0}^{\infty }\left[ \frac{1}{\tau }%
\int_{0}^{\tau }H\left( s\right) ds\right] \left[ \tau f\left( \tau \right) %
\right] d\tau  \\ 
\ \\ 
=\int_{0}^{\infty }\left[ \int_{0}^{\tau }H\left( s\right) ds\right] f\left(
\tau \right) d\tau =\int_{0}^{\infty }H\left( s\right) \left[
\int_{s}^{\infty }f\left( \tau \right) d\tau \right] ds \\ 
\ \\ 
=\int_{0}^{\infty }H\left( s\right) \bar{F}\left( s\right)
ds=\int_{0}^{\infty }\frac{f\left( s\right) }{\bar{F}\left( s\right) }\bar{F}%
\left( s\right) ds \\ 
\ \\ 
=\int_{0}^{\infty }f\left( s\right) ds=1.
\end{array}%
\right.   \label{A262}
\end{equation}%
Substituting Eq. (\ref{A262}) into the right-hand side of Eq. (\ref{A261})
yields Eq. (\ref{73}):%
\begin{equation}
\int_{0}^{\infty }\left[ \bar{H}\left( \tau \right) -H\left( \infty \right) %
\right] f_{1}\left( \tau \right) d\tau =\frac{1}{\mu }-H\left( \infty
\right) .  \label{A263}
\end{equation}

\subsection{Optimization calculations}

Evidently, the average $\bar{H}\left( \tau \right) =\frac{1}{\tau }\int_{0}^{\tau }H\left( t\right)$ is a function of the timer $0<\tau<\infty$. Differentiating the average with respect to the timer yields
\begin{equation}
\left. 
\begin{array}{l}
\bar{H}^{\prime }\left( \tau \right) =\frac{H\left( \tau \right) \tau
-\int_{0}^{\tau }H\left( t\right) dt}{\tau ^{2}} \\ 
\ \\ 
=\frac{1}{\tau }\left[ H\left( \tau \right) -\bar{H}\left( \tau \right) %
\right] .
\end{array}%
\right.  \label{A301}
\end{equation}
Differentiating Eq. (\ref{A301}) with respect to the timer further yields 
\begin{equation}
\left. 
\begin{array}{l}
\bar{H}^{\prime \prime }\left( \tau \right) =\frac{\left[ H^{\prime }\left(
\tau \right) -\bar{H}^{\prime }\left( \tau \right) \right] \tau -\left[
H\left( \tau \right) -\bar{H}\left( \tau \right) \right] }{\tau ^{2}} \\ 
\ \\ 
=\frac{1}{\tau }\left[ H^{\prime }\left( \tau \right) -2\bar{H}^{\prime
}\left( \tau \right) \right] .
\end{array}%
\right.  \label{A303}
\end{equation}%
Hence, at critical timers -- $\tau_{c}$ that satisfy $\bar{H}^{\prime}(\tau_{c})=0$ -- we have: 
\begin{equation}
\bar{H}^{\prime \prime }(\tau_{c})=\frac{1}{\tau_{c} }H^{\prime }(\tau_{c}).
\end{equation}

Last, consider the negative logarithmic derivative of the input's density function, $G\left( t\right) =-f^{\prime }\left(
t\right) /f\left( t\right) $. Using the function $G(t)$, the derivative of the input's hazard function admits the following representation:
\begin{equation}
\left. 
\begin{array}{l}
H^{\prime }\left( t\right) =\left[ \frac{f\left( t\right) }{\bar{F}\left(
t\right) }\right] ^{\prime } \\ 
\ \\ 
=\frac{f^{\prime }\left( t\right) \bar{F}\left( t\right) -f\left( t\right) %
\left[ -f\left( t\right) \right] }{\bar{F}\left( t\right) ^{2}} \\ 
\ \\ 
=\frac{f\left( t\right) }{\bar{F}\left( t\right) }\frac{f^{\prime }\left(
t\right) }{f\left( t\right) }+\left[ \frac{f\left( t\right) }{\bar{F}\left(
t\right) }\right] ^{2} \\ 
\ \\ 
=H\left( t\right) \left[ -G\left( t\right) \right] +H\left( t\right) ^{2} \\ 
\ \\ 
=H\left( t\right) \left[ H\left( t\right) -G\left( t\right) \right].
\end{array}%
\right.  \label{A305}
\end{equation}

\end{document}